\documentclass[eqsecnum,showpacs,pre,twocolumn]{revtex4}%
\usepackage{graphicx}
\usepackage{epsfig}
\usepackage{dcolumn}
\usepackage{amsfonts}
\usepackage{amsmath}
\usepackage{amssymb}
\usepackage{bm}%
\begin{document}
\title{Transport properties of directed percolation clusters at the upper critical dimension}
\author{Olaf Stenull}
\affiliation{Fachbereich Physik, Universit\"at Duisburg-Essen, Campus Essen, 45117 Essen,
Germany }
\author{Hans-Karl Janssen}
\affiliation{Institut f\"{u}r Theoretische Physik III, Heinrich-Heine-Universit\"{a}t,
40225 D\"{u}sseldorf, Germany}
\date{\today}

\begin{abstract}
\noindent We study the transport properties of directed percolation clusters
at the upper critical dimension $d_{c} = 4+1$, where critical fluctuations
induce logarithmic corrections to the leading (mean-field) scaling behavior.
Employing field theory and renormalization group methods we calculate these
logarithmic corrections up to and including the next to leading correction for
a variety of observables, viz.\ the connectivity, i.e., the probability that
two given points are connected, the average two-point resistance and some of
the fractal masses describing percolation clusters. Furthermore, we study
logarithmic corrections for the multifractal moments of the current
distribution on directed percolation clusters.

\end{abstract}
\maketitle

\newcommand{\brm}[1]{\bm{{\rm #1}}}

\section{Introduction}

\label{intro} 
\begin{figure}[ptb]
\centerline{\includegraphics[width=5.5cm]{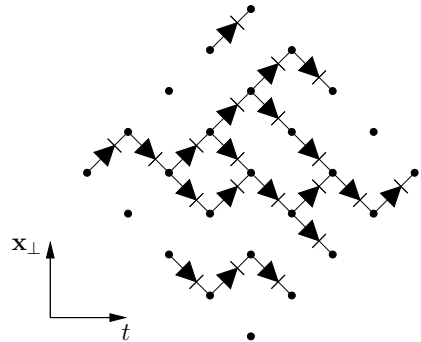}}\caption{(1+1)-dimensional
directed bond percolation realized as a ransom diode network on a tilted
square lattice. Electrical current can percolate only along the distinguished
direction $t$.}%
\label{diodenetworknetwork}%
\end{figure}

\noindent Directed percolation (DP)~\cite{Hi01,JaTa04} is
perhaps the simplest model for directed connectivity in disordered systems. It
differs from conventional isotropic percolation (IP)~\cite{reviewsIP} in that
activity (in the following electric current) can percolate only along a
certain distinguished direction. A simple and intuitive realization of DP is
the random diode network (RDN), where nearest neighboring bonds on a tilted
hypercubic lattice are randomly occupied (with probability $p$) with diodes,
see Fig.~\ref{diodenetworknetwork}. Evidently, the effective conductivity
along the distinguished direction depends on the occupation probability $p$.
If $p$ is small, all cluster of connected sites are finite so that the
conductivity vanishes in the limit of large distances. On the other hand, if
$p$ exceeds a certain critical threshold $p_{c}$, there is a finite
probability to find a directed path between two points, leading to a finite
resistance when averaged over many independent samples. At the critical
threshold $p=p_{c}$, the system undergoes a continuous phase transition where
clusters of conducting paths display a fractal structure. In contrast to IP,
these clusters are anisotropic and at criticality they are rather self-affine
than self-similar.
\begin{figure}[ptbptb]
\centerline{\includegraphics[width=8.6cm]{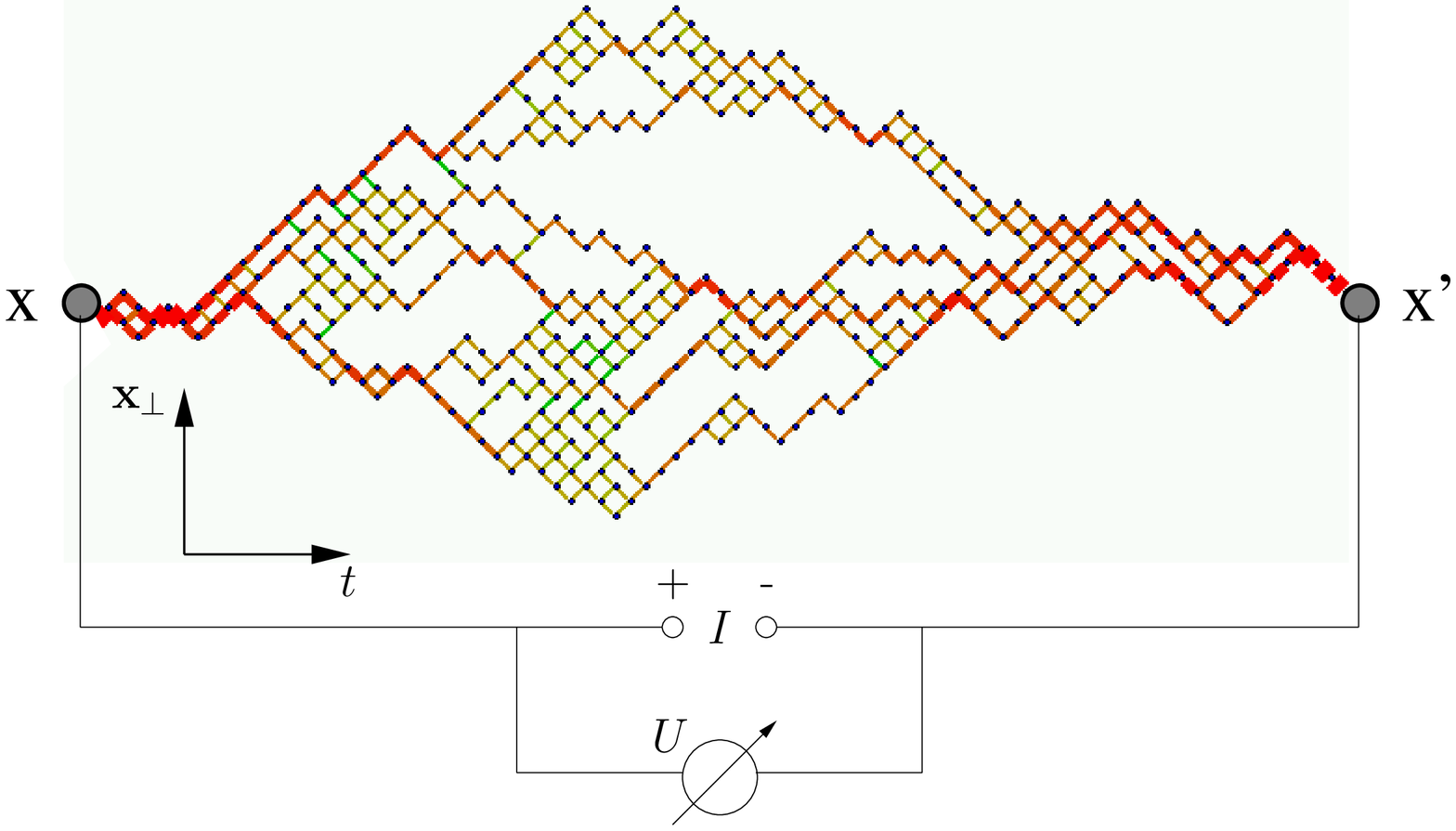}}\caption{Two-port setup for
measuring the resistance of a DP cluster. The figure shows a generic
current-carrying backbone when a current $I$ is inserted at a point
$\brm{x}$ and withdrawn at a point $\brm{x}^{\prime}$. The thickness of
the bonds indicates the intensity of the current. The thickest bonds are the
so-called red bonds that carry the full current.}%
\label{network}%
\end{figure}

In this paper we are interested in the electrical transport properties of DP
clusters as realized by the RDN. Due to the self-affinity of the critical
clusters, transport quantities like the average resistance $M_{R} (\brm{x},
\brm{x}^{\prime})$ between two connected points $\brm{x}$ and
$\brm{x}^{\prime}$ exhibit a scaling behavior as it is typical for
(anisotropic) critical phenomena. Below the upper critical dimension $d_{c} =
4+1$ this scaling behavior is characterized by critical exponents which are,
due to the effect of fluctuations, anomalous. Above $d_{c}$ fluctuations are
unimportant and the scaling behavior is purely of mean-field type. Right at
$d_{c}$, fluctuations lead to logarithmic corrections to the mean-field
behavior. To date, the critical exponents for variety of transport quantities
characterizing DP clusters are known to second order in an $\varepsilon$
expansion about $d_{c}$, namely the exponents for $M_{R} (\brm{x},
\brm{x}^{\prime})$%
~\cite{janssen_stenull_directedLetter_2000,stenull_janssen_rdnResistance_2001}%
, for the fractal masses $M_{B} (\brm{x}, \brm{x}^{\prime})$,
$M_{\text{red}} (\brm{x}, \brm{x}^{\prime})$ and $M_{\text{min}}
(\brm{x}, \brm{x}^{\prime})$ respectively of the backbone, the red
(singly connected) bonds and the chemical (shortest)
path~\cite{janssen_stenull_directedLetter_2000,stenull_janssen_nonlinearRDN_2001}
as well as the exponents for the multifractal moments $M_{I}^{(l)}
(\brm{x}, \brm{x}^{\prime})$ of the current
distribution~\cite{stenull_janssen_epl_2001,stenull_janssen_pre_2002_RDNmultifrac}
(cf.\ Fig.~\ref{network}). Logarithmic corrections to these quantities have
not been studied hitherto. With the computer resources and elaborate
algorithms available today, computer simulations can explore high dimensions
and it seems to be within reach to investigate the transport properties of DP
clusters at $d_{c}$ numerically with good precision. Thus, we feel that it
becomes important to investigate the corresponding logarithmic corrections analytically.

The system in statistical physics, for which logarithmic corrections have been
studied perhaps most thoroughly, are linear polymers. The work of Sch\"{a}fer
and coworkers~\cite{GHS94,GHS99} showed that knowing the leading logarithmic
corrections is, at least in the case of linear polymers, not sufficient to
obtain a satisfactory agreement between theory and simulations. Rather, one
has to push the theory beyond the leading corrections, which involves, in
addition to the basic analysis of the RG mapping, the calculation of scaling
functions. We expect that higher-order logarithmic corrections will be
likewise important for the transport properties of DP clusters. In part, this
expectation is corroborated by the experience one has with logarithmic
corrections in other percolation problems. A comparison between
numerical~\cite{grassberger_private} and
analytical~\cite{janssen_stenull_dip_log} results for dynamic isotropic
percolation at the respective upper critical dimension clearly indicates the
importance of the higher order logarithmic corrections.
Simulations~\cite{luebeck_willmann_2004} on dynamical aspects of DP at $d_{c}%
$, on the other hand, show a satisfactory agreement with our theoretic
predictions~\cite{janssen_stenull_dp_log} already at the level of the leading
logarithmic correction. There are no simulations available yet for a
comparison to our analytic results on logarithmic corrections for transport on
IP clusters~\cite{stenull_janssen_rrn_log}.

In this paper we investigate the logarithmic corrections for transport on DP
clusters up to and including the next to leading logarithmic correction. For
simplicity, we assume that $\brm{x}=(\brm{0},0)$ and $\brm{x}%
^{\prime}=(\brm{0},t)$. We calculate the quantities of interest as
functions of the longitudinal or time-like variable $t$, that is $M_{R}(t)$,
$M_{B}(t)$, $M_{\text{red}}(t)$ and $M_{\text{min}}(t)$ as well as
$M_{I}^{(l)}(t)$. By and large as a byproduct, we also obtain the
connectivity, i.e., the probability that the points $\brm{x}=(\brm{0}%
,0)$ and $\brm{x}^{\prime}=(\brm{0},t)$ are connected, $P(t)$.

The outline of this paper is as follows. In Sec.~\ref{theModel} we briefly review
a generalization of the RDN, the random resistor diode network (RRDN), that
certain advantages over the RDN with respect to setting up a field theoretic
model. Then, we explain our field theoretic model, including its variants, and
highlight its physical contents. Section~\ref{renormalization} reviews
renormalization group results for the RRDN in order to establish notation and
to provide known results that we need as an input as we go along.
Section~\ref{logarithmicCorrections} represents the main part of this paper.
Here we derive the desired logarithmic corrections and we state our final
results. Concluding remarks are given in Sec.~\ref{conclusions}. Technical
details on the calculation of scaling functions are relegated to an Appendix.

\section{The model, its variants and the physical contents}

\label{theModel}

\subsection{The random resistor diode network}

\label{RRDN} 

The RRDN introduced by Redner~\cite{red_81&82a,red_83,perc} is a
simple model for electric transport in irregular media that comprises both, DP
and IP. A RRDN consists of a $d$-dimensional hyper-cubic lattice in which
nearest-neighbor sites are connected by a resistor, a positive diode
(conducting only in a preferred direction), a negative diode (conducting only
opposite to the preferred direction), or an insulator with respective
probabilities $p$, $p_{+}$, $p_{-}$, and $q=1-p-p_{+}-p_{-}$. To be specific,
let us choose $\brm{n}=1/\sqrt{d}\left(  1,\dots,1\right)  $ as the
preferred direction and assume that the bonds $\underline{b}_{\langle
i,j\rangle}$ between two nearest neighboring lattice sites $i$ and $j$ are
directed so that $\underline{b}_{\langle i,j\rangle}\cdot\brm{n}>0$.
Moreover, let us suppose that the directed bonds obey the generalized Ohm's
law
\begin{align}
\sigma_{\underline{b}_{\langle i,j\rangle}}\big(V_{\underline{b}_{\langle
i,j\rangle}}\big)\,V_{\underline{b}_{\langle i,j\rangle}}=I_{\underline
{b}_{\langle i,j\rangle}}\,, \label{avoid}%
\end{align}
where $V_{\underline{b}_{\langle i,j\rangle}}=V_{j}-V_{i}$ is the voltage drop
over the bond between sites $j$ and $i$ and $I_{\underline{b}_{\langle
i,j\rangle}}$ is the associated current. The bond conductances $\sigma
_{\underline{b}}=\sigma\,\gamma_{\underline{b}}$ are random variables where
$\gamma_{\underline{b}}$ takes on the values $1$, $\theta\left(  V\right)  $,
$\theta\left(  -V\right)  $ (as usual, $\theta$ denotes the Heaviside
function), and $0$ with respective probabilities $p$, $p_{+}$, $p_{-}$, and
$q$, and where $\sigma$ is a positive constant. Note that this assumption
means that the diodes are idealized, i.e., under forward-bias voltage they
behave as \textquotedblleft ohmic\textquotedblright\ resistors whereas they
are insulating under backward-bias voltage. Further below we will also deal
with nonlinear voltage-current characteristics under forward-bias of the type
$V\sim I^{r}$. The three dimensional phase diagram, i.e., the tetrahedron
spanned by the four probabilities, features a non-percolating and three
percolating phases, viz.\ isotropic, positively directed, and negatively
directed, with continuous transitions between the four phases. For a detailed
discussion of the phase diagram see Refs.~\cite{red_81&82a,red_83}.

The long-length-scale behavior of the critical RDN and that of the RRDN at
either of the 2 transitions from the non-percolating to the directed
percolating phases are equivalent, provided of course, that the distinguished
direction of the RDN corresponds to the distinguished direction of the RRDN or
its opposite direction, respectively. Thus, we may investigate the transport
properties of DP clusters as realized by the RDN via studying the RRDN at
either of the transition the 2 transitions from the non-percolating to the
directed percolating phases. We prefer to work with the RRDN because it has
certain advantages with respect to setting up a field theoretic model. For
details on deriving a field theoretic model for the RRDN we refer to
Refs.~\cite{janssen_stenull_directedLetter_2000,stenull_janssen_rdnResistance_2001}%
. In the vicinity of the transitions from the non-percolating to either of the
directed percolating phases this model can be written in the form of a dynamic
response functional~\cite{janssen_dynamic,deDominicis&co,janssen_92},
\begin{align}
{\mathcal{J}}=  &  \int d^{d_{\perp}}r\,dt\bigg\{\frac{1}{2}\sum_{\vec
{\lambda}\neq\vec{0}}s_{-\vec{\lambda}}\Big[\rho\left(  \tau-\nabla^{2}%
+w\vec{\lambda}^{2}\right) \nonumber\label{dynFktnal}\\
&  +\left[  \theta\left(  \lambda_{0}\right)  -\theta\left(  -\lambda
_{0}\right)  \right]  \frac{\partial}{\partial t}\Big]s_{\vec{\lambda}%
}\nonumber\\
&  +\frac{\rho g}{6}\sum_{\vec{\lambda},\vec{\lambda}^{\prime},\vec{\lambda
}+\vec{\lambda}^{\prime}\neq\vec{0}}s_{-\vec{\lambda}}s_{-\vec{\lambda
}^{\prime}}s_{\vec{\lambda}+\vec{\lambda}^{\prime}}\bigg\}\,,
\end{align}
whose ingredients have a meaning as follows. $s_{\vec{\lambda}}=s_{\vec
{\lambda}}(\brm{x})$ is an order parameter field which lives on the
$d$-dimensional real space with coordinates $\brm{x}=(\brm{r},t)$, where
$t\sim x_{\Vert} = \brm{x}\cdot\brm{n}$ is the longitudinal part of
$\brm{x}$ along the preferred direction and $\brm{r}=\brm{x}_{\bot}$
is the corresponding ($d_{\perp}=d-1$)-dimensional transversal part. In
addition to the space coordinate, the order parameter field depends on a
$D$-fold replicated current variable $\vec{\lambda}$ living on a
$D$-dimensional replica space which we will explain a little further below.
Regarding the current variable the order parameter field satisfies the
constraint $s_{\vec{0}}(\brm{x})=0$. The parameter $\tau$ specifies the
distance in phase space to the transition from the non-percolating to the
directed phase of interest which occurs in mean field theory at $\tau=0$. $w$
is proportional to $\sigma^{-1}$ and $\rho$ is a kinetic coefficient. The
meaning of $\lambda_{0}$ will become clear shortly when we elaborate on the
replica space.

For regularization purposes, the replica space resembles a discretized
$D$-dimensional torus, i.~e., $\vec{\lambda}=\vec{k}\Delta\lambda$ where
$\vec{k}$ is a $D$-dimensional vector with integer components $k^{(\alpha)}$
satisfying $-M<k^{(\alpha)}\leq M$ and $k^{(\alpha)}=k^{(\alpha)}\mod (2M)$.
To extract physical quantities from the replica formulation one has to study
the limit $D\rightarrow0$, $M\rightarrow\infty$ with $(2M)^{D}\rightarrow1$
and $\Delta\lambda=\lambda_{M}/\sqrt{M}\rightarrow0$%
~\cite{footnote_redundancy}. In addition to these settings dictated by
regularization issues, we work near the limit when all the components of
$\vec{\lambda}$ are equal and continue to large imaginary values, i.e., we
set~\cite{harris_87}
\begin{align}
\lambda^{(\alpha)}=i\lambda_{0}+\xi^{(\alpha)} \label{lambdaChoice}%
\end{align}
with real $\lambda_{0}$ and $\xi^{(\alpha)}$ satisfying $|\lambda_{0}|\gg1$
and $\sum_{\alpha=1}^{D}\xi^{(\alpha)}=0$. This specific choice for the
replica current is tailored so that: First, we can assign a sign (positive or
negative direction) to the multidimensional replica currents. Second, it
allows us in actual calculations involving summations or integrations over the
replica currents to resort to the saddle point approximation which is crucial
since the elementary circuit elements as modelled by Eq.~(\ref{avoid}) are
non-linear. Finally, we invoke the conditions $\lambda_{0}^{2}\ll D^{-1}$ and
$D\lambda_{0}^{2}\ll\vec{\xi}^{2}\ll1$ making $\vec{\lambda}^{2}=\vec{\xi}%
^{2}-D\lambda_{0}^{2}$ a small positive quantity so that expansions in powers
of $\vec{\lambda}^{2}$ are reliable.

After this excursion into formal aspects which we feel, however, was necessary
for a proper definition of the model, we now turn to its physical contents.
One of the basic quantities describing the electric transport properties of DP
clusters is the average macroscopic resistance
\begin{align}
M_{R}(\brm{x},\brm{x}^{\prime})=\frac{\langle\chi_{+}(\brm{x}%
,\brm{x}^{\prime})R_{+}(\brm{x},\brm{x}^{\prime})\rangle_{C}%
}{P(\brm{x},\brm{x}^{\prime})} \label{defAvRes}%
\end{align}
when an external current $I$ is inserted at a point $\brm{x}$ and extracted
at another point $\brm{x}^{\prime}$ provided, that the 2 points are
positively connected. In Eq.~(\ref{defAvRes}), $R_{+}(\brm{x}%
,\brm{x}^{\prime})$ is the total resistance if $I$ is inserted at
$\brm{x}$ and extracted at point $\brm{x}^{\prime}$ and $\chi
_{+}(\brm{x},\brm{x}^{\prime})$ is an indicator function that gives the
value 1 one if $\brm{x}$ and $\brm{x}^{\prime}$ are positively connected
(if $I$ can percolated from $\brm{x}$ to $\brm{x}^{\prime}$), and zero
otherwise. $\langle\cdots\rangle_{C}$ denotes the disorder average over all
configurations of the diluted lattice and $P(\brm{x},\brm{x}^{\prime
})=\langle\chi_{+}(\brm{x},\brm{x}^{\prime})\rangle_{C}$ is the
connectivity that measures the probability for $\brm{x}$ and $\brm{x}%
^{\prime}$ being positively connected. The model is set up so that the 2-point
correlation function
\begin{align}
G_{2}(\brm{x},\brm{x}^{\prime},\vec{\lambda})=\left\langle
s(\brm{x},\vec{\lambda})s(\brm{x}^{\prime},-\vec{\lambda})\right\rangle
\label{defCorr}%
\end{align}
is a generating function for $M_{R}(x,x^{\prime})$. This can be seen by noting
that
\begin{align}
G_{2}(\brm{x},\brm{x}^{\prime},\vec{\lambda})  &  =\left\langle
\exp\left[  -\frac{\vec{\lambda}^{2}}{2}R_{+}(\brm{x},\brm{x}^{\prime
})\right]  \right\rangle _{C}\nonumber\\
&  =P(\brm{x},\brm{x}^{\prime})\left\{  1-\frac{\vec{\lambda}^{2}}%
{2}M_{R}(\brm{x},\brm{x}^{\prime})+\cdots\right\}  , \label{struct}%
\end{align}
up to an unimportant multiplicative factor that goes to 1 in the replica limit
$D\rightarrow0$. Thus, once we have calculated the 2-point correlation
function, as we are able by using renormalized field theory, we can extract
the average resistance by using
\begin{align}
M_{R}(\brm{x},\brm{x}^{\prime})=\frac{\partial}{\partial(-\vec{\lambda
}^{2}/2)}\ln G_{2}\left(  \brm{x},\brm{x}^{\prime},\vec{\lambda}\right)
\Big|_{\vec{\lambda}=0}. \label{exploitGenFkt}%
\end{align}

\subsection{The nonlinear RRDN}

\label{nonlinearRRDN} By generalizing the model we can study much more than
just the average resistance. Suppose that the directed bonds obey the
non-linear Ohm's law~\cite{kenkel_straley_82}
\begin{align}
\sigma_{\underline{b}_{\langle i,j\rangle}}\big(V_{\underline{b}_{\langle
i,j\rangle}}\big)\,V_{\underline{b}_{\langle i,j\rangle}}\,\big|V_{\underline
{b}_{\langle i,j\rangle}}\big|^{s-1}=I_{\underline{b}_{\langle i,j\rangle}}\,.
\end{align}
The dynamic functional of the nonlinear RRDN constituted by these elements is
of the same form as the functional~(\ref{dynFktnal}), however, with the
replacement~\cite{janssen_stenull_directedLetter_2000,stenull_janssen_nonlinearRDN_2001}
\begin{align}
w\vec{\lambda}^{2}\rightarrow w_{r}\Lambda_{r}(\vec{\lambda})\,,
\end{align}
where $r=1/s$ and
\begin{align}
\Lambda_{r}(\vec{\lambda})=-\sum_{\alpha=1}^{D}\left(  -i\lambda^{(\alpha
)}\right)  ^{r+1}\,.
\end{align}
The 2-point correlation function in the generalized model has the property
that, up to an unimportant constant,
\begin{align}
G_{2}(\brm{x},\brm{x}^{\prime},\vec{\lambda})  &  =\left\langle
\exp\left[  -\frac{\Lambda_{r}(\vec{\lambda})}{r+1}R_{r,+}(\brm{x}%
,\brm{x}^{\prime})\right]  \right\rangle _{C}\,\nonumber\\
&  =P(\brm{x},\brm{x}^{\prime})\left\{  1-\frac{\Lambda_{r}(\vec
{\lambda})}{r+1}M_{R_{r}}(\brm{x},\brm{x}^{\prime})+\cdots\right\}  ,
\label{structR}%
\end{align}
where $R_{r,+}(\brm{x},\brm{x}^{\prime})$ is the nonlinear resistance
between the two terminals. Thus, the 2-point correlation function is here a
generating function for the average nonlinear resistance
\[
M_{R_{r}}(\brm{x},\brm{x}^{\prime})=\frac{\langle\chi_{+}(\brm{x}%
,\brm{x}^{\prime})R_{r,+}(\brm{x},\brm{x}^{\prime})\rangle_{C}%
}{P(\brm{x},\brm{x}^{\prime})}\,,
\]
which can be calculated by using
\begin{align}
M_{R_{r}}(\brm{x},\brm{x}^{\prime})=\frac{\partial}{\partial
(-\Lambda_{r}(\vec{\lambda})/(r+1))}\ln G_{2}\left(  \brm{x},\brm{x}%
^{\prime},\vec{\lambda}\right)  \Big|_{\vec{\lambda}=0}.
\label{exploitGenFktnl}%
\end{align}
The generalized RRDN has the benefit that it features the free parameter $r$
and that this parameter can be used to access several physical quantities. For
$r\rightarrow1$ one retrieves, of course, the linear average resistance
$M_{R}(\brm{x},\brm{x}^{\prime})$. For $r\rightarrow-1^{+}$, one obtains
the mass (the average number of bonds) $M_{B}$ of the backbone,
\begin{align}
M_{B}(\brm{x},\brm{x}^{\prime})\sim\lim_{r\rightarrow-1^{+}}M_{R_{r}%
}(\brm{x},\brm{x}^{\prime})\,,
\end{align}
as one can see straightforwardly by considering the overall dissipated
electric power. Following the lines of Blumenfeld and
Aharony~\cite{blumenfeld_aharony_85}, one can show that
\begin{align}
M_{\text{red}}(\brm{x},\brm{x}^{\prime})\sim\lim_{r\rightarrow\infty
}M_{R_{r}}(\brm{x},\brm{x}^{\prime})\,,
\end{align}
for the mass of the red bonds and
\begin{align}
M_{\text{min}}(\brm{x},\brm{x}^{\prime})\sim\lim_{r\rightarrow0^{+}%
}M_{R_{r}}(\brm{x},\brm{x}^{\prime})\,,
\end{align}
for the mass of the chemical path.

\subsection{The noisy RRDN}

\label{noisyRRDN} To study multifractal aspects of transport on DP clusters we
can generalize the RRDN so that it features static noise. Suppose that the
directed bonds have a current-voltage characteristic as stated in
Eq.~(\ref{avoid}) but that the value of the conductance of occupied bonds
under forward bias fluctuates statically about some average value. To model
this effect we set $\sigma_{\underline{b}}=\varsigma_{\underline{b}}%
\,\gamma_{\underline{b}}$ where $\varsigma_{\underline{b}}$ is a random
variable distributed according to some distribution function $f$ with mean
$\overline{\varsigma}=\sigma$ and higher cumulants $\Delta^{(l\geq2)}$
satisfying $\Delta^{(l)}\ll\sigma^{l}$. The condition on the cumulants is
imposed to suppress unphysical negative conductances. The distribution
function $f$ might for example be Gaussian, however, our considerations are
not limited to this particular choice.

In order to access the multifractal properties of this noisy RRDN we must
treat the average $\langle\cdots\rangle_{C}$ over the configurations of the
diluted lattice and the noise average $\{ \cdots \}_{f}$
independently. For this purpose we go beyond the usual replica trick and use a
second replication parameter so that the replicated currents
$\tensor{\lambda}$ are $(D\times E)$ tuples and not merely $D$
tuples~\cite{park_harris_lubensky_87}. Though the replica space for this model
is somewhat more complicated than the one discussed in Sec.~\ref{RRDN},
essentially the same regularization issues appear here and we can make much
the same settings and impose much the same conditions as in Sec.~\ref{RRDN}
albeit in a somewhat more general form. For details we refer to
Refs.~\cite{stenull_janssen_epl_2001,stenull_janssen_pre_2002_RDNmultifrac}
where we derived and analyzed a field theoretic model for the noisy RRDN. The
dynamic functional embodying this model near the transitions from the
nonpercolating to the directed percolating phases has the same form as the
functional~(\ref{dynFktnal}). However, we have to make the
replacement~\cite{stenull_janssen_epl_2001,stenull_janssen_pre_2002_RDNmultifrac}
\begin{align}
\vec{\lambda}\rightarrow\tensor{\lambda}
\end{align}
and there is an additional term in the dynamic functional,
\begin{align}
\mathcal{J}\rightarrow\mathcal{J}+\sum_{l=2}^{\infty}v_{l}\mathcal{V}^{(l)}\,,
\end{align}
with dangerously irrelevant interactions
\begin{align}
\mathcal{V}^{(l)}=\frac{\rho}{2}\int d^{d_{\perp}}r\,dt\,\mathcal{O}%
^{(l)}(\brm{r},t)\
\end{align}
with coupling constants $v_{l}\sim\Delta^{(l)}/\sigma^{2l}$. Here, the
$\mathcal{O}^{(l)}$ are composite fields
\begin{align}
\mathcal{O}^{(l)}={\sum_{\tensor{\lambda}}}s_{-\tensor{\lambda}}%
K_{l}\big(\tensor{\lambda}\big)s_{\tensor{\lambda}}%
\end{align}
featuring the homogeneous polynomials
\begin{align}
K_{l}(\tensor{\lambda})=\sum_{\beta=1}^{E}\bigg[\sum_{\alpha=1}^{D}\left(
\lambda^{(\alpha,\beta)}\right)  ^{2}\bigg]^{l}\,.
\end{align}

The two-point correlation function $G_{2}(\brm{x},\brm{x}^{\prime
},\tensor{\lambda})$ of the noisy RRDN contains a lot of information on the
transport properties of DP clusters. In particular it contains the so-called
noise cumulants
\begin{align}
C_{R}^{(l)}(\brm{x},\brm{x}^{\prime})=\frac{\big\langle\chi
_{+}(\brm{x},\brm{x}^{\prime})\big\{R_{+}(\brm{x},\brm{x}^{\prime
})^{l}\big\}_{f}^{(c)}\big\rangle_{C}}{P(\brm{x},\brm{x}^{\prime})}\,,
\end{align}
where $\{R_{+}(\brm{x},\brm{x}^{\prime})^{l}\}_{f}^{(c)}$ stands for the
$l$th cumulant of the resistance $R_{+}(\brm{x},\brm{x}^{\prime})$ with
respect to $f$. This can be understood by noting that, up to an unimportant
multiplicative constant that goes to one in the replica limit $D\rightarrow
0$,
\begin{align}
&  G_{2}(\brm{x},\brm{x}^{\prime},\tensor{\lambda})\nonumber\\
&  =\left\langle \exp\left[  \sum_{l=1}^{\infty}\frac{(-1/2)^{l}}{l!}%
K_{l}(\tensor{\lambda})\big\{R_{+}\left(  \brm{x},\brm{x}^{\prime
}\right)  ^{l}\big\}_{f}^{(c)}\right]  \right\rangle _{C}\nonumber\\
&  =P(\brm{x},\brm{x}^{\prime})\Bigg\{1+\sum_{l=1}^{\infty}%
\frac{(-1/2)^{l}}{l!}K_{l}(\tensor{\lambda})\,C_{R}^{(l)}(\brm{x}%
,\brm{x}^{\prime})+\cdots\Bigg\}\,, \label{structM}%
\end{align}
i.e., the two-point correlation function is a generating function for the
noise cumulants which can be extracted by using
\begin{align}
C_{R}^{(l)}(\brm{x},\brm{x}^{\prime})=\frac{\partial}{\partial\left[
\frac{(-1/2)^{l}}{l!}K_{l}(\tensor{\lambda})\right]  }\ln G_{2}(\brm{x}%
,\brm{x}^{\prime},\tensor{\lambda})\bigg|_{\tensor{\lambda}=\tensor{0}}\,.
\label{exploitGenFktM}%
\end{align}
Although the noise cumulants are interesting in their own right, we are
primarily interested in a family of observables that is more intuitive,
viz.\ the multifractal moments
\begin{align}
M_{I}^{(l)}(\brm{x},\brm{x}^{\prime})=\frac{\left\langle \chi
_{+}(\brm{x},\brm{x}^{\prime})\sum_{\underline{b}}\left(  I_{\underline
{b}}/I\right)  ^{2l}\right\rangle _{C}}{P(\brm{x},\brm{x}^{\prime})}%
\end{align}
of the current distribution. Using Cohn's theorem~\cite{cohn_50}, one can show
that
\begin{align}
M_{I}^{(l)}(\brm{x},\brm{x}^{\prime})\sim C_{R}^{(l)}(\brm{x}%
,\brm{x}^{\prime})\,. \label{M-C}%
\end{align}
This relation allows us to study the multifractal current distribution
indirectly via the noise cumulants, which is important, since, to our
knowledge, there is no direct way of calculating the moments of the current
distribution by field theoretic means.

\section{Renormalization}

\label{renormalization} In recent years we have investigated the scaling
behavior of the average resistance, the fractal masses, and the multifractal
moments of the current distribution below the upper critical dimension
$d_{\perp}=4$%
~\cite{janssen_stenull_directedLetter_2000,stenull_janssen_rdnResistance_2001,stenull_janssen_nonlinearRDN_2001,stenull_janssen_epl_2001,stenull_janssen_pre_2002_RDNmultifrac}%
. In particular we calculated various scaling exponents and fractal dimensions
by using renormalized field theory. The aim of this section is to briefly
review central elements of our field theory to provide background, to
establish notation and to gather previous results that we will need as input
as we go on. For background on renormalized field theory in general we refer
to~\cite{amit_zinn-justin}.

\subsection{The linear and the nonlinear RRDN}

To keep our presentation compact, we will review here only the general case,
i.e., the nonlinear RRDN. The linear RRDN can be retrieved by simply letting
$r \to1$.

As usual, a central building block of our renormalization group analysis is a
diagrammatic perturbation theory. Here, the diagrammatic elements for
constructing Feynman diagrams are the vertex $\rho g$ and the propagator
\begin{align}
\widetilde{G}(\brm{p},t,\vec{\lambda})=\widetilde{G}_{+}(\brm{p}%
,t,\vec{\lambda})+\widetilde{G}_{-}(\brm{p},t,\vec{\lambda})\,,
\end{align}
with
\begin{align}
&  \widetilde{G}_{\pm}(\brm{p},t,\vec{\lambda})=\theta\left(  \pm t\right)
\theta\left(  \pm\lambda_{0}\right) \nonumber\label{defGpm}\\
&  \times\exp\left[  \mp\rho\left(  \tau+\brm{p}^{2}+w_{r}\Lambda_{r}%
(\vec{\lambda})\right)  t\right]  \left(  1-\delta_{\vec{\lambda},\vec{0}%
}\right)  \, ,
\end{align}
where the factor $(1-\delta_{\vec{\lambda},\vec{0}})$ enforces the constraint
$\vec{\lambda}\neq\vec{0}$. For the actual calculations it is sufficient to
keep either $\widetilde{G}_{+}(\brm{p},t,\vec{\lambda})$ or $\widetilde
{G}_{-}(\brm{p},t,\vec{\lambda})$ and we choose to keep $\widetilde{G}%
_{+}(\brm{p},t,\vec{\lambda})$.

Calculating the conducting Feynman diagrams in dimensional regularization, one
encounters ultraviolet divergences in the form of poles in the deviation
$\varepsilon=4-d_{\perp}$ from the upper critical transversal dimension
$d_{\perp}=4$. These $\varepsilon$ poles can be handled by using the
renormalization scheme
\begin{subequations}
\label{renorScheme}%
\begin{align}
&  s\rightarrow{\mathring{s}}=Z^{1/2}s\,,\qquad\;\rho\rightarrow
{\mathring{\rho}}=Z^{-1}Z_{\rho}\rho\,,\\
&  \tau\rightarrow{\mathring{\tau}}=Z_{\rho}^{-1}Z_{\tau}\tau\,,\quad
w_{r}\rightarrow\mathring{w}{_{r}}=Z_{\rho}^{-1}Z_{w_{r}}w_{r}\,,\\
&  g^{2}\rightarrow{\mathring{g}}^{2}=Z^{-1}Z_{\rho}^{-2}Z_{u}G_{\varepsilon
}^{-1}u\mu^{\varepsilon}\,,
\end{align}
where $\mu^{-1}$ is the usual arbitrary mesoscopic length scale. The factor
$G_{\varepsilon}=(4\pi)^{-d_{\perp}/2}\Gamma(1+\varepsilon/2)$, with $\Gamma$
denoting the Gamma function, is introduced for convenience. $Z$, $Z_{\tau}$,
$Z_{\rho}$, and $Z_{u}$ are the usual DP $Z$ factors known to second order in
$\varepsilon$~\cite{janssen_81,janssen_2000}. In our work on the linear RRDN,
see
Refs.~\cite{janssen_stenull_directedLetter_2000,stenull_janssen_rdnResistance_2001}%
, we determined $Z_{w}=Z_{w_{1}}$ to second order in $\varepsilon$. Studying
the nonlinear
RRDN~\cite{janssen_stenull_directedLetter_2000,stenull_janssen_nonlinearRDN_2001}%
, we showed to two-loop order that
\end{subequations}
\begin{subequations}
\begin{align}
Z_{0}  &  =\lim_{r\rightarrow0}Z_{w_{r}}=Z\,,\\
Z_{-1}  &  =\lim_{r\rightarrow-1^{+}}Z_{w_{r}}=1\,,\\
Z_{\infty}  &  =\lim_{r\rightarrow\infty}Z_{w_{r}}=Z_{\tau}\,.
\end{align}
It is not unreasonable to expect that these special relations hold to
arbitrary order in perturbation theory, i.e., that they are Ward identities
resulting from the symmetries of the system. Identifying these symmetries and
proving the presumable Ward identities is a challenging open issue for future
work. Here, for our calculations to follow below we need to know the
renormalization factors explicitly only to one-loop order,
\end{subequations}
\begin{subequations}
\label{renorFactors}%
\begin{align}
&  Z=1+\frac{u}{4\,\varepsilon}\,,\quad Z_{\rho}=1+\frac{u}{8\,\varepsilon
}\,,\\
&  Z_{\tau}=1+\frac{u}{2\,\varepsilon}\,,\quad Z_{u}=1+\frac{2\,u}%
{\varepsilon}\,,\\
&  Z_{w_{r}}=1+\frac{u}{2\,\varepsilon}\Big(1-\frac{1}{2^{r+1}}\Big)\,.
\end{align}

The fact that the unrenormalized theory must be independent of the inverse
length scale introduced in the renormalization process can be used in a
routine fashion to set up a Gell-Mann--Low renormalization group equation
(RGE) for the correlation functions. For an $N$ point function this RGE reads
\end{subequations}
\begin{align}
\left[  \mathcal{D}_{\mu,r}+\frac{N}{2}\gamma\right]  G_{N}\left(  \left\{
\brm{r},\rho t,w_{r}\Lambda_{r}(\vec{\lambda})\right\}  ;\tau,u,\mu\right)
=0\,.
\end{align}
where
\begin{align}
\mathcal{D}_{\mu,r}=\mu\frac{\partial}{\partial\mu}+\beta\frac{\partial
}{\partial u}+\tau\kappa\frac{\partial}{\partial\tau}+w_{r}\zeta_{r}%
\frac{\partial}{\partial w_{r}}+\rho\zeta_{\rho}\frac{\partial}{\partial\rho
}\,.
\end{align}

The functions featured in the RGE are given to two-loop order by
\begin{subequations}
\label{betauRes}%
\begin{align}
\beta(u)  &  =-\varepsilon u+\frac{3u^{2}}{2}-\Big(169+106\ln\frac{4}%
{3}\Big)\frac{u^{3}}{128}+O\left(  u^{4}\right)  \,,\label{wilsonZetaR}\\
\kappa(u)  &  =\frac{3u}{8}-\,\Big(7+10\ln\frac{4}{3}\Big)\frac{7u^{2}}%
{256}+O\left(  u^{3}\right)  \,,\\
\gamma(u)  &  =-\frac{u}{4}+\Big(6-9\ln\frac{4}{3}\Big)\frac{u^{2}}%
{32}+O\left(  u^{3}\right)  \,,\\
\zeta_{\rho}(u)  &  =-\frac{u}{8}+\,\Big(17-2\ln\frac{4}{3}\Big)\frac{u^{2}%
}{256}+O\left(  u^{3}\right)  \,,\\
\zeta_{r}(u)  &  =\zeta_{r,1}\,u+\zeta_{r,2}\,u^{2}+O\left(  u^{3}\right)  \,.
\end{align}
The coefficient of the first order term in Eq.~(\ref{wilsonZetaR}) is known
for arbitrary $r$,
\end{subequations}
\[
\zeta_{r,1}=\Big(\frac{3}{8}-\frac{1}{2^{r+2}}\Big)
\]
The coefficient $\zeta_{r,2}$ is known only for particular values of $r$. Our
two-loop analysis of the linear RRDN with $r=1$ gave
\begin{align}
\zeta_{1,2}=-\frac{5}{32}\,.
\end{align}
For the remaining values of $r$ of interest here, $\zeta_{r,2}$ can be
inferred readily from the relations
\begin{subequations}
\begin{align}
\zeta_{-1}(u)  &  =\gamma(u)-\zeta_{\rho}(u)\,,\quad\zeta_{0}(u)=-\zeta_{\rho
}(u)\,,\\
\zeta_{\infty}(u)  &  =\kappa(u)\,.
\end{align}
stemming from the special relations between the $Z$-factors.

In the following we will use for the RGE-functions an abbreviated notation of
the type $f(u)=f_{1}u+f_{2}u^{2}+\cdots$. For example, we will write
Eq.~(\ref{wilsonZetaR}) as $\beta(u)=\beta_{1}u+\beta_{2}u^{2}+\beta_{3}%
u^{3}+O(u^{4})$ and likewise for the other RGE-functions.

\subsection{The noisy RRDN}

The parameters $v_{l}$ featured in the dynamic functional of the noisy RRDN
are, in contrast to the relevant parameter $w=w_{1}$, dangerously irrelevant,
i.e., they are irrelevant on dimensional grounds but they must not be
neglected in studying the noise cumulants because we otherwise inevitably
loose the information we are interested in. Due to their irrelevance, the
$v_{l}$ cannot be treated in the same fashion as the relevant $w$. A
proper treatment of the $v_{l}$ can be achieved by looking at insertions of
the dangerously irrelevant interactions $\mathcal{V}^{(l)}$ into Feynman
diagrams. Due to the irrelevance, insertions of $\mathcal{V}^{(l)}$ generate a
multitude of terms corresponding to interactions with equal or lower naive
dimension than $\mathcal{V}^{(l)}$. All these interactions have to be taken
into account in the renormalization process. The interactions of lower naive
dimension, however, merely lead to subdominant corrections and can be ignored
for our purposes. Keeping all the interactions of the same naive dimension, we
have a renormalization in matrix form
\end{subequations}
\begin{align}
\underline{\mathcal{V}}^{(l)}\rightarrow\underline{\mathcal{\mathring{V}}%
}^{(l)}=\left(  \underline{\underline{Z}}^{(l)}\right)  ^{-1}\underline
{\mathcal{V}}^{(l)}\,
\end{align}
where $\underline{\mathcal{V}}^{(l)}=(\mathcal{V}^{(l)},\mathcal{V}_{2}%
^{(l)},\cdots)$ is a vector that contains all the interactions generated by
$\mathcal{V}^{(l)}$ by the renormalization process including $\mathcal{V}%
^{(l)}$ itself. The $\mathcal{V}^{(l)}$ are distinguished by the feature that
the interactions generated by $\mathcal{V}^{(l)}$ do not in turn generate
$\mathcal{V}^{(l)}$, or as we say, that the corresponding composite fields
$\mathcal{O}^{(l)}$ are master
operators~\cite{stenull_janssen_epl_2000,stenull_janssen_2001,stenull_janssen_epl_2001,stenull_janssen_pre_2002_RDNmultifrac}%
. Because of their subordinate role, we refer to the remaining operators
$\mathcal{O}_{2}^{(l)}$ and so on as servants. The master operators are
associated with renormalization matrices $\underline{\underline{Z}}%
^{(l)}=\underline{\underline{1}}+O(u)$, where $\underline{\underline{1}}$
stands for the unit matrix, of a particularly simple structure,
\begin{align}
\underline{\underline{Z}}^{(l)}=\left(
\begin{array}
[c]{cccc}%
Z^{(l)} & \ast & \cdots & \ast\\
0 & \ast & \cdots & \ast\\
\vdots & \vdots & \ddots & \vdots\\
0 & \ast & \cdots & \ast
\end{array}
\right)  ,
\end{align}
where the $\ast$ symbolize arbitrary elements. As a consequence of the simple
structure of $\underline{\underline{Z}}^{(l)}$ the servants can be neglected
in calculating the scaling index of their master. This can be seen as follows.
Due to Eq.~(\ref{exploitGenFktM}) we are ultimately interested in derivatives
of the Green's function with insertions of $\mathcal{V}^{(l)}$,
$G_{2;\mathcal{V}^{(l)}}$, with respect to $K_{l}(\tensor{\lambda})$
evaluated at $\tensor{\lambda}=0$. From the renormalization of the master
interaction,
\begin{align}
\mathcal{V}^{(l)}=Z^{(l)}\mathcal{\mathring{V}}^{(l)}+\sum_{\alpha\geq
2}Y_{\alpha}^{(l)}\mathcal{\mathring{V}}_{\alpha}^{(l)}\,,
\end{align}
where the $Y_{\alpha}^{(l)}$ are elements of $\underline{\underline{Z}}^{(l)}%
$, it follows that
\begin{align}
G_{2}(\brm{r},t)_{\mathcal{V}^{(l)}}=Z^{(l)}G_{2}(\brm{r}%
,t)_{\mathcal{\mathring{V}}^{(l)}}+\sum_{\alpha\geq2}Y_{\alpha}^{(l)}%
G_{2}(\brm{r},t)_{\mathcal{\mathring{V}}_{\alpha}^{(l)}}\mathcal{\,}.
\label{pandur}%
\end{align}
The coefficients $Y_{\alpha}^{(l)}$ pertaining to the servants are required to
make $G_{2}(\brm{r},t)_{\mathcal{V}^{(l)}}$ free of $\varepsilon$-poles.
However, only the first term on the right hand side of Eq.~(\ref{pandur})
gives a nonzero contribution when we differentiate with respect to
$K_{l}(\tensor{\lambda})$ and then set $\tensor{\lambda}=0$ since only the
insertion of $\mathcal{\mathring{V}}^{(l)}$ produces the polynomial structure
of $K_{l}(\tensor{\lambda})$. Hence, as long as we restrict ourselves to the
properties of the noise cumulants $C_{R}^{(l)}(\brm{x},\brm{x}^{\prime
})$ we only need for our practical purposes the multiplicative
renormalizations of the $\mathcal{V}^{(l)}$, i.e., we only need the element
$Z^{(l)}$. We can set all the other renormalizations $Y_{\alpha}^{(l)}$
pertaining to the servants formally to zero.

Understanding that the renormalization of the dangerously irrelevant
interactions $\mathcal{V}^{(l)}$ is subtle but that we can ignore these
subtleties for our practical purposes, we treat their couplings $v_{l}$ in
much the same way as $w$. Doing so we have to bear in mind, of course, that this
procedure only makes sense if we expand all Feynman diagrams in powers of
$v_{l}$ and truncate this expansion after linear order (which corresponds to
using single insertions). Then, we renormalize the $v_{l}$ by setting
\begin{align}
v_{l} \to\mathring{v}_{l}=Z_{\rho}^{-1}Z_{v_{l}}v_{l}=Z^{(l)}v_{l}
\label{renorScheme2}%
\end{align}
$Z_{v_{l}}$ is known for arbitrary $l=0,1,2,\cdots$ to one-loop order and for
most important values of $l$ to two-loop order. Below we will need to know
$Z_{v_{l}}$ explicitly to one-loop order for calculating the desired
logarithmic corrections. To this order $Z_{v_{l}}$ is related to the
$Z_{w_{r}}$ by
\begin{align}
Z_{v_{l}}=Z_{w_{2l-1}}=1+\left(  1-\frac{1}{4^{l}}\right)  \frac
{u}{2\varepsilon} \label{renorFactors2}%
\end{align}
as we show in the appendix. The two-point correlation function is now governed
by the RGE
\begin{align}
&  \left\{  \left[  \mathcal{D}_{\mu,1}+\gamma\right]  +\sum_{l}\gamma
^{(l)}v_{l}\frac{\partial}{\partial v_{l}}\right\} \nonumber\\
&  \times G_{2}\left(  \brm{r},\rho t,w\tensor{\lambda}^{2};\{v_{l}%
K_{l}(\tensor{\lambda})\},\tau,u,\mu\right)  =0\,.
\end{align}
The Gell-Mann--Low function $\gamma^{(l)}$ stemming from $Z^{(l)}$ is%
\begin{align}
\gamma^{(l)}=\gamma_{1}^{(l)}u+\gamma_{2}^{(l)}u^{2}+O(u^{3})\,,
\label{gamma-l}%
\end{align}
where
\begin{align}
\gamma_{1}^{(l)}=\zeta_{2l-1,1}=\frac{3}{8}-\frac{1}{2^{2l+1}}\,.
\end{align}
$\gamma^{(l)}$ satisfies the relations $\gamma^{(0)}=\zeta
_{-1}=\gamma-\zeta_{\rho}$ and $\gamma^{(1)}=\zeta_{1}$. $\gamma_{2}^{(l)}$ is
stated for $l=2,\cdots,5$ in Table~\ref{gamma2table}.

\begin{table}[ptb]
\caption{The coefficients $\gamma_{2}^{(l)}$ appearing in Eq.~(\ref{gamma-l}%
).}%
\label{gamma2table}%
\begin{tabular}
[c]{c||c|c|c|c}\hline\hline
$\quad l \quad$ & $2$ & $3$ & $4$ & $5$\\\hline
$\gamma_{2}^{(l)}$ & $-0.24249$ & $-0.26523$ & $-0.26988$ & $-0.27045$%
\\\hline\hline
\end{tabular}
\end{table}
\medskip

\section{Logarithmic corrections}

\label{logarithmicCorrections}

The RGE can be solved by the method of characteristics whereby one introduces
a single flow parameter $\ell$ and sets up characteristic equations that
describe how the scaling parameters transform under a change of $\ell$. The
characteristic for the momentum scale $\mu$ is particularly simple and has the
solution $\bar{\mu}(\ell)=\mu\ell$, i.e., a change of $\ell$ corresponds to a
change of the external inverse length scale. With help of the solution to the
remaining characteristics one obtains \begin{widetext}%
\begin{align}
\label{GrFuSkal}
G_{2}\left(  \brm{r},\rho t,w_{r}\Lambda_{r}(\lambda);\{v_{l}K_{l}%
(\lambda)\},\tau,u,\mu\right)  =(\mu\ell)^{d_{\perp}}\bar{Z}(\ell)%
G_{2}\left(  \mu\ell\,\brm{r},(\mu\ell)^{2}\bar{\rho}(\ell)\,t,\frac
{\bar{w}_{r}(\ell)\,\Lambda_{r}(\lambda)}{(\mu\ell)^{2}};\left\{  \frac
{\bar{v}_{l}(\ell)\,K_{l}(\lambda)}{(\mu\ell)^{2}}\right\}  ,\frac{\bar{\tau
}(\ell)}{(\mu\ell)^{2}},\bar{u}(\ell),1\right)  .
\end{align}
\end{widetext}
as a solution to the RGE. This formula applies to the nonlinear as well as to
the noisy RRDN with the understanding that we have to set $v_{l} =0$ and
$\lambda=\vec{\lambda}$ in the former and $r=1$ and $\lambda=\tensor{\lambda}$
in the latter case. At this stage the scaling solution~(\ref{GrFuSkal}) is
still rather formal since $\bar{Z}(\ell)$, $\bar{\rho}(\ell)$, $\bar{\tau
}(\ell)$, $\bar{w}_{r}(\ell)$, $\bar{v}(\ell)$ and $\bar{u}(\ell)$ require
specification. Below the upper critical dimension, these quantities display
power law behavior for $\ell\rightarrow0$ described by the well known critical
exponents of the DP universality class, the resistance exponents $\phi_{r}$,
and the multifractal exponents $\psi_{l}$ of the RRDN. Directly in $d_{\perp
}=4$, they depend logarithmically on $\ell$ and hence their behavior is
qualitatively different from the lower dimensional case.

Now we will state and solve the characteristics directly for $d_{\perp}=4$.
The characteristic for the dimensionless coupling constant $u$ is given by
\begin{align}
\ell\frac{d\bar{u}}{d\ell}=\beta(\bar{u})\,. \label{Char-u}%
\end{align}
The solution to this differential equation for $\varepsilon=0$ is
\begin{align}
\ell=\ell(\bar{u})=\ell_{0}\bar{u}^{-\beta_{3}/\beta_{2}^{2}}\exp
\bigg[-\frac{1}{\beta_{2}\bar{u}}+O(\bar{u})\bigg]\,, \label{l(w)}%
\end{align}
where $\ell_{0}$ is an integration constant. The remaining characteristics are
all of the same structure, namely
\begin{align}
\ell\frac{d\ln\bar{Q}(\bar{u})}{d\ell}=q(\bar{u})\,, \label{genericChar}%
\end{align}
where $Q$ is a placeholder for $Z$, $\rho$, $\tau$, $w_{r}$, and $v_{l}$,
respectively, and $q$ is placeholder for $\gamma$, $\zeta_{\rho}$, $\kappa$,
$\zeta_{r}$, and $\gamma^{(l)}$, respectively. Exploiting $\ell\,d/d\ell=\beta
d/d\bar{u}$ one obtains the solution
\begin{align}
\bar{Q}(\bar{u})=Q_{0}\,\bar{u}^{q_{1}/\beta_{2}}\exp\bigg[\frac{(q_{2}%
\beta_{2}-q_{1}\beta_{3})}{\beta_{2}^{2}}\bar{u}+O(\bar{u}^{2})\bigg],
\label{Q(w)}%
\end{align}
where $Q_{0}$ is a non universal integration constant.

After having reviewed some of the cornerstones of the field theory of the RRDN
we will now determine the critical behavior of the connectivity, the average
resistance, the fractal masses of the backbone, the chemical path and the red
bonds as well as the multifractal moments of the current distribution at the
upper critical dimension $d_{\perp}=4$. For simplicity, we set in the
following $\brm{x}^{\prime}=(\brm{r}^{\prime},0)=(\brm{0},0)$ and
$\brm{x}=(\brm{r},0)=(\brm{0},t)$ and restrict our attention to the
behavior of the aforementioned quantities as functions of the time-like
variable $t$. To this end we choose
\begin{align}
(\mu\ell)^{2}\bar{\rho}(\ell)\,t=X_{0}\,, \label{Wahl_X}%
\end{align}
where $X_{0}$ is a constant of order 1. With this choice $\bar{u}$ and $\ell$
tend to zero for $\rho\mu^{2}t\rightarrow\infty$. From Eq.~(\ref{l(w)}) and
Eq.~(\ref{Q(w)}), specialized to $\bar{\rho}$, we obtain with $\beta_{2}=3/2$
\begin{align}
t=t_{0}\,\bar{u}^{-2a/3}\exp\left(  \frac{4}{3\bar{u}}\right)  \left[
1+O(\bar{u})\right]  \label{tBehave}%
\end{align}
where $t_{0}=X_{0}/(\mu^{2}\ell_{0}^{2}\,\rho_{0})$ is yet another non
universal constant and $a$ is given by
\begin{align}
a=\frac{\beta_{2}\zeta_{\rho,1}-2\beta_{3}}{2\beta_{2}}=\frac{157}{192}%
+\frac{53}{96}\ln\frac{4}{3}=0.97653\,.
\end{align}
At certain stages it will be more convenient to use, instead of using the
original $t$, the variable
\begin{align}
s=\frac{3}{4}\ln\bigl(t/t_{0}\bigr)\,. \label{s(t)}%
\end{align}
For $s$, Eq.~(\ref{tBehave}) translates into
\begin{align}
s=\bar{u}^{-1}-a\ln\bar{u}+O(\bar{u})\,. \label{t(w)}%
\end{align}
Finally, we find by using Eq.~(\ref{t(w)})
\begin{align}
\bar{u}=s^{-1}\exp\bigg[a\frac{\ln s}{s}+O\Big(\frac{\ln^{2}s}{s^{2}}%
,\frac{\ln s}{s^{2}},\frac{1}{s^{2}}\Big)\bigg] \label{w(s)}%
\end{align}
for the dimensionless coupling constant as a function of the longitudinal
coordinate $s$.

At this point we know the general scaling form~(\ref{GrFuSkal}) of the
$2$-point correlation function and thus we have an important part of the
information that we need to understand the critical behavior of connectivity,
the fractal masses and the multifractal moments at the upper critical
dimension. However, for determining the logarithmic corrections to the scaling
behavior of these quantities beyond the leading correction, we need to know
some additional information about the scaling function appearing on the right
hand side of Eq.~(\ref{GrFuSkal}).

Expanding the right hand side of Eq.~(\ref{GrFuSkal}) at criticality $\tau=0$
to linear order in $\bar{w}_{r}\Lambda_{r}$ and $\bar{v}_{l}K_{l}$, and by
using Eq.~(\ref{Wahl_X}) we find that the two-point correlation function is at
$d_{\perp}=4$ of the form \begin{widetext}%
\begin{align}
&  G_{2}\big(\brm{0},\rho t,w_{r}\Lambda_{r}(\lambda);\{v_{l}K_{l}%
(\lambda)\},0,u,\mu\big)=(\mu\ell)^{4}\bar{Z}(\bar{u})G_{2}\big(\brm{0}%
,X_{0},0;\{0\},0,\bar{u},1\big)\nonumber\\
&  \times\bigg\{1+(\mu\ell)^{-2}\bar{w}_{r}(\bar{u})\Lambda_{r}(\lambda
)\,g_{2}^{\prime}\big(X_{0},\bar{u}\big)+(\mu\ell)^{-2}\sum_{l}\bar{v}%
_{l}(\bar{u})K_{l}(\lambda)\,g_{2}^{(l)}\big(X_{0},\bar{u}\big)\cdots
\bigg\}\,,\label{generalscal}%
\end{align}
\end{widetext}
with expansion coefficients $g_{2}^{\prime}(X_{0},\bar{u})=\left.  \partial\ln
G_{2}/\partial w_{r}\Lambda_{r}\right\vert _{\lambda=0}$ and $g_{2}%
^{(l)}(X_{0},\bar{u})=\left.  \partial\ln G_{2}/\partial v_{l}K_{l}\right\vert
_{\lambda=0}$. For the second logarithmic correction we have to calculate the
functions $G_{2}(\brm{0},X_{0},0;\{0\},0,\bar{u},1)$, $g_{2}^{\prime}%
(X_{0},\bar{u})$ and $g_{2}^{(l)}(X_{0},\bar{u})$ to one-loop order. Some
details on these calculations can be found in the Appendix.

\subsection{Connectivity, average resistance and fractal masses}

Our one-loop calculation sketched in the Appendix yields that the scaling
functions relevant for the connectivity, the average resistance and the
fractal masses are given by
\begin{subequations}
\label{resScalFuB}%
\begin{align}
G_{2}\big(\brm{0},X_{0},0;\bar{u},0,1\big)  &  = c_{0}\bigl[1+A_{P}%
\,\bar{u}+O(\bar{u}^{2})\bigr]\,,\label{AmpP}\\
g_{2}^{\prime}\big(X_{0},\bar{u}\big)  &  = c_{1}\bigl[1+A_{r}\,\bar{u}%
+O(\bar{u}^{2})\bigr]\,, \label{AmpR}%
\end{align}
where $c_{0}$ and $c_{1}$ are nonuniversal constants and where $A_{P}$ and
$A_{r}$ are amplitudes that can depend on $X_{0}$. $A_{P}$ turns out being
independent of $X_{0}$,
\end{subequations}
\begin{align}
A_{P}=\frac{3}{16}\,, \label{AP}%
\end{align}
For the amplitude $A_{r}$ we find
\begin{align}
A_{r}=\frac{1-\ln2-\mathcal{Z}}{8}\Big(1-\frac{1}{2^{r}}\Big)\, , \label{Ar}%
\end{align}
where $\mathcal{Z=Z}(X_{0})$ is a nonuniversal constant that we have defined,
to stay consistent with our work on logarithmic corrections for dynamic
properties of DP \cite{janssen_stenull_dp_log}, as $\mathcal{Z}(X_{0}%
)=C_{E}+\ln( 2X_{0})$ with $C_{E}=0.577215...$ being Euler's constant.
Alternatively, it is possible to choose the arbitrary parameter $X_{0}$ in
such a way that $\mathcal{Z}=1-\ln2$ so that the amplitude $A_{r}$ becomes
simply zero. Physically, this would only change the nonuniversal time scale
$t_{0}$ entering the scaling functions through Eq.~(\ref{tBehave}). However,
we prefer to choose $\mathcal{Z}$ as we did for two reasons. First, as
mentioned above, this allows us to stay consistent with previous work. Second,
leaving $\mathcal{Z}$ in our formulas has the advantage that $\mathcal{Z}$ can
be used as a fit parameter when it comes to comparing our results to simulations.

Now we are finally in the position to assemble our results for the
connectivity and the average (nonlinear) resistance. Merging
Eqs.~(\ref{structR}), (\ref{generalscal}), and (\ref{AmpP}) we obtain for the
connectivity
\begin{align}
P(t)\sim(\mu\ell)^{4}\,\bar{Z}(\bar{u})\bigl[1+A_{P}\,\bar{u}+O(\bar{u}%
^{2})\bigr]\,. \label{structP}%
\end{align}
Exploiting our choice~(\ref{Wahl_X}) and using our knowledge about the
solutions of the characteristics we find
\begin{subequations}
\begin{align}
t^{2}\,P(t)/P_{0}  &  =\left(1+\frac{3\bar{u}}{16}\right)\exp\left(
-c_{P}\,\bar{u}\right)  \,\left[  1+O\left(  \bar{u}^{2}\right)  \right]
\label{resPa}\\
&  =\left(1+\frac{3}{16s}\right)\bigg\{1-\frac{c_{P}}{s}+O\left(  \frac{\ln^{2}%
s}{s^{2}},\frac{\ln s}{s^{2}},\frac{1}{s^{2}}\right)  \bigg\} \label{resPb1}%
\end{align}
where $P_{0}$ is a nonuniversal constant and
\end{subequations}
\begin{align}
c_{P}=\frac{2\,\zeta_{\rho,2}-\gamma_{2}}{\beta_{2}}=-\frac{1}{192}%
\Big(7-34\ln\frac{4}{3}\Big)=0.01448\,.
\end{align}
Equation~(\ref{resPa}) in conjunction with Eq.~(\ref{tBehave}) can be viewed
as a parametric representation of the tuple $(t^{2}P(t),t)$ with $\bar{u}$
serving as the free parameter. Equation~(\ref{resPb1}) states our result in a
more traditional form. Though perhaps less intuitive, the parametric form has
the conceptual advantage that it involves only one expansion variable,
viz.~the effective coupling constant $\bar{u}$ whereas in the traditional form
functions of the longitudinal variable such as $1/s^{2}$, $\ln s/s^{2}$,
$\ln^{2}s/s^{2}$ and so on compete against each other. Note that the usually
leading logarithmic correction is absent in case of the connectivity $P(t)$.
Hence, one has to go to the next order, as we did, to see a deviation from the
mean-field field behavior $P(t)\propto t^{-2}$.

Next we turn to the average nonlinear resistance. Using Eqs.~(\ref{struct}),
(\ref{GrFuSkal}), (\ref{generalscal}), (\ref{AmpR}) as well as
Eq.~(\ref{exploitGenFktnl}) yields
\begin{align}
M_{R_{r}}(t)\sim(\mu\ell)^{-2}\,\bar{w}_{r}(\bar{u})\bigl[1+A_{r}\,\bar
{u}+O(\bar{u}^{2})\bigr]\,.
\end{align}
Inserting several intermediate results we are led to
\begin{subequations}
\label{resR}%
\begin{align}
&  t^{-1}\,M_{R_{r}}(t)/M_{R_{r},0}=\bigl(\bar{u}^{-1}+B\bigr)^{a_{r}}%
\exp\left(  -c_{r}\,\bar{u}\right) \nonumber\\
&  \qquad\qquad\qquad\qquad\qquad\qquad\times\left[  1+O\left(  \bar{u}%
^{2}\right)  \right] \label{resRa}\\
&  =\bigl(s+B\bigr)^{a_{r}}\bigg\{1-\frac{b_{r}\ln s+c_{r}}{s}+O\left(
\frac{\ln^{2}s}{s^{2}},\frac{\ln s}{s^{2}},\frac{1}{s^{2}}\right)
\bigg\} \label{resRb}%
\end{align}
with a nonuniversal constant $M_{R_{r},0}$ and
\end{subequations}
\begin{subequations}
\label{ParR}%
\begin{align}
a_{r}  &  =-\frac{\zeta_{\rho,1}+\zeta_{r,1}}{\beta_{2}}=-\frac{1-2^{-r}}%
{6}\,,\\
b_{r}  &  =a\,a_{r}\,,\\
c_{r}  &  =\frac{(\zeta_{\rho,1}+\zeta_{r,1})\beta_{3}}{\beta_{2}^{2}}%
-\frac{\zeta_{\rho,2}+\zeta_{r,2}}{\beta_{2}}\,,\\
B  &  =\frac{A_{r}}{a_{r}}=-\frac{3}{4}\bigl(1-\ln2-\mathcal{Z}\bigr)\,.
\end{align}
As explained above, the physically most important limits of $r$ are
$r\rightarrow1,-1^{+},\infty,0^{+}$ since these limits yield respectively the
average linear resistance and the fractal masses of the backbone, the red
bonds, and the chemical path. The numerical values of $b_{r}$ and $c_{r}$ in
the limits $r\rightarrow1,-1^{+},\infty$ are collected in
Table~\ref{table:nonlinExpAmp}. In the limit $r\rightarrow0^{+}$ the
quantities $a_{r}$, $b_{r}$, $c_{r}$, and $A_{r}$ vanish and the amplitude $B$
does not make much sense. This vanishing of the parameters had to be expected
because in DP the length of the chemical path between the two points
$(\brm{0},0)$ and $(\brm{0},t)$ is essentially $t$ and that hence
\end{subequations}
\begin{align}
M_{\text{min}}\propto t \label{scalingMin}%
\end{align}
with no logarithmic correction. The fact that our result is in conformity with
this anticipated behavior is reassuring and we rate it as an important
consistency check for our calculation.
\begin{table}[ptb]
\caption{Values of the numbers $b_{r}$ and $c_{r}$ appearing in
Eqs.~(\ref{resR}).}%
\label{table:nonlinExpAmp}%
\begin{tabular}
[c]{c||c|c|c}\hline\hline
$\quad r \quad$ & $-1$ & $1$ & $\infty$\\\hline
$b_{r} $ & $0.16275$ & $-0.08137$ & $-0.16275$\\\hline
$c_{r} $ & $0.10211$ & $-0.02519$ & $-0.03589$\\\hline\hline
\end{tabular}
\end{table}

\subsection{Multifractal moments}

The logarithmic corrections for the multifractal moments can be calculated in
much the same way as the corrections for the average nonlinear resistance. Our
one-loop calculation sketched in the Appendix yields
\begin{align}
g_{2}^{(l)}(X_{0},\bar{u})=c_{1}\bigl[1+A_{2l-1}\,\bar{u}+O(\bar{u}%
^{2})\bigr]\,, \label{AmpM}%
\end{align}
where the amplitude $A_{2l-1}$ can be read off from Eq.~(\ref{Ar}) setting
$r=2l-1$. Using Eqs.~(\ref{structM}), (\ref{M-C}), (\ref{generalscal}),
(\ref{AmpM}) as well as (\ref{exploitGenFktM} and Eq.~(\ref{AmpM}) we obtain
\begin{align}
M_{I}^{(l)}(t)\sim(\mu\ell)^{-2}\bar{v}^{(l)}(\bar{u})\,\bigl(1+A_{2l-1}%
\,\bar{u}+O(\bar{u}^{2})\bigr)\,.
\end{align}
Recalling our choice for the flow parameter, Eq.~(\ref{Wahl_X}), and using
that the solution of the characteristic for $\bar{v}^{(l)}$ is of the
form~(\ref{Q(w)}) we obtain
\begin{subequations}
\label{resM}%
\begin{align}
&  t^{-1}\,M_{I}^{(l)}(t)/M_{I,0}^{(l)}=\bigl(\bar{u}^{-1}+B\bigr)^{a^{(l)}%
}\exp(-c^{(l)}\,\bar{u})\,\left[  1+O\left(  \bar{u}^{2}\right)  \right] \\
&  =\bigl(s+B\bigr)^{a^{(l)}}\bigg\{1-\frac{b^{(l)}\ln s+c^{(l)}}{s}+O\left(
\frac{\ln^{2}s}{s^{2}},\frac{\ln s}{s^{2}},\frac{1}{s^{2}}\right)  \bigg\}
\end{align}
where $M_{I,0}^{(l)}$ are nonuniversal constants. The amplitude $B$ is the
same as the one in Eq.~(\ref{ParR}). The parameters $a^{(l)}$ and $b^{(l)}$
are related to the parameters of Eqs.~(\ref{ParR}) by%
\end{subequations}
\begin{align}
a^{(l)}=a_{2l-1}\,,\qquad b^{(l)}=b_{2l-1}\,,
\end{align}
and $c^{(l)}$ is given by%
\begin{align}
c_{I}^{(l)}=\frac{(\zeta_{\rho,1}+\gamma_{1}^{(l)})\beta_{3}}{\beta_{2}^{2}%
}-\frac{\zeta_{\rho,2}+\gamma_{2}^{(l)}}{\beta_{2}}\,.
\end{align}
Numerical values for the coefficient $c_{I}^{(l)}$ can be found in
Table~\ref{table:noisyExpAmp}.
\begin{table}[ptb]
\caption{Values of the numbers $b^{(l)}$ and $c^{(l)}$, appearing in
Eqs.~(\ref{resM}).}%
\label{table:noisyExpAmp}
\begin{tabular}
[c]{c||c|c|c|c}\hline\hline
$\quad l \quad$ & $2$ & $3$ & $4$ & $5$\\\hline
$b^{(l)} $ & $-0.14241$ & $-0.15766$ & $-0.16148$ & $-0.16243$\\\hline
$c^{(l)} $ & $-0.03263$ & $-0.03371$ & $-0.03466$ & $-0.03530$\\\hline\hline
\end{tabular}
\end{table}
Note that $M_{I}^{(1)}\sim C_{R}^{(1)}=M_{R}$ as it should. Note also that $M_{I}^{(0)}\sim
C_{R}^{(0)}=M_{B}$. This must hold since $\sum_{b}(I_{b}/I)^{2l}$ coincides
for $l\rightarrow0$ with the number of current-carrying bonds and thus
$M_{I}^{(0)}=M_{B}$.

\section{Concluding remarks}

\label{conclusions}In summary, we have studied the electrical transport
properties of DP clusters by using the methods of renormalized field theory.
To our knowledge, logarithmic corrections for the connectivity, the average
resistance, the fractal masses of the backbone, the red bonds, the chemical
path and the multifractal moments of the current distribution have not been
considered hitherto. We calculated these corrections up to and including the
next to leading correction.

We hope that our work triggers complementary numerical simulations. With
todays computer hardware and sophisticated algorithms, our results should be
testable by numerical work. Because we went beyond just calculating the
leading corrections, we are optimistic that our results to compare well with
simulations, perhaps even quantitatively.


\appendix*

\section{Amplitudes}

In this Appendix we deliver some details on our calculation of the amplitudes
entering the second logarithmic corrections. First, let us look at the
two-point correlation function at zero-loop level. As a function of the
longitudinal and the transversal coordinates we have
\begin{align}
G_{2}^{(0)}(\brm{r},t,\lambda)=\int_{\brm{p}}\exp(i\brm{p}%
\cdot\brm{r})\,\widetilde{G}_{+}(\brm{p},t,\lambda)\,,
\end{align}
where $\int_{\brm{p}}$ is an abbreviation for $1/(2\pi)^{d_{\perp}}\int
d^{d_{\perp}}p$. To treat the nonlinear and the noisy RRDN in one go, we use
\begin{align}
\widetilde{G}_{+}(\brm{p},t,\lambda)=\bigl(1-\delta_{\lambda,0}%
\bigr)\exp\Big[-\rho\bigl(\tau+\brm{p}^{2}+L(\lambda)\bigr)t\Big]\
\end{align}
as our Gaussian propagator. Here, $L(\lambda)$ stands for the polynomial
\begin{align}
L(\lambda)=w_{r}\Lambda_{r}(\lambda)+\sum_{l}v_{l}K_{l}(\lambda)\,.
\end{align}
As in Sec.~\ref{logarithmicCorrections} it is understood that we have to set
$v_{l}=0$ and $\lambda=\vec{\lambda}$ or $r=1$ and $\lambda=\tensor{\lambda}$
to retrieve the nonlinear or the noisy RRDN, respectively.

To one-loop order the correlation function is governed by the Dyson-equation%
\begin{align}
&  G_{2}(\brm{r},t,\lambda)-G_{2}^{(0)}(\brm{r},t,\lambda)\nonumber\\
&  =\int_{\brm{p}}\exp(i\brm{p}\cdot\brm{r})\int_{0}^{t}dt_{1}%
\int_{0}^{t_{1}}dt_{2}\widetilde{G}(\brm{p},t-t_{1},\lambda)\nonumber\\
&  \times\Sigma(\brm{p},t_{1}-t_{2},\lambda)\widetilde{G}(\brm{p}%
,t_{2},\lambda)\,, \label{Dyson}%
\end{align}
\begin{figure}[ptb]
\centerline{\includegraphics[width=4cm]{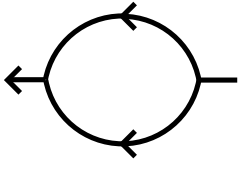}}
\caption{Diagrammatic representation of the self-energy $\Sigma(\brm{q}%
,t,\lambda)$ at one-loop order.}%
\label{decomposition}%
\end{figure}
where $\Sigma$ stands for the self-energy $\Sigma$ depicted in
Fig.~\ref{decomposition},
\begin{align}
&  \Sigma(\brm{p},s,\lambda)=\frac{\rho^{2}g^{2}}{2}\sum_{\kappa}%
\int_{\brm{q}}\biggl\{ -2\exp\bigl(-\rho L(\lambda)s\bigr)\nonumber\\
&  +\sum_{\kappa}\exp\Big[-\rho\bigl(L(\lambda/2+\kappa)+L(\lambda
/2-\kappa)\bigr)s\Big]\biggr\}\nonumber\\
&  \times\int_{\brm{q}}\exp\Big[-\rho\bigl(2\tau+(\brm{p}/2+\brm{q)}%
^{2}+(\brm{p}/2-\brm{q)}^{2}\bigr)s\Big]\,.
\end{align}
Note that we have included here an extra term independent of $\kappa$ to switch from the initially restricted current summation to a current summation without the restriction $\kappa \neq 0$. The
integration over $\brm{q}$ is straigthforward. After switching from the current summation to an  integration over $\kappa$ we employ the saddle point method which we can since $\lambda_{0}\gg0$. From the symmetry of the integrand it follows irrespectively of the detailed form of
$L(\lambda)$ that the locus of the saddle-point is at $\kappa=0$.
Consequently, in the limit $D\rightarrow0$,%
\begin{align}
&  \Sigma(\brm{p},s,\lambda)=\frac{\rho^{2}g^{2}}{2}\biggl(\exp\bigl(-2\rho
L(\lambda/2)s\bigr)-2\exp\bigl(-\rho L(\lambda
)s\bigr)\biggr)\nonumber\label{hunger}\\
&  \times\bigl(8\pi\rho s\bigr)^{-d_{\bot}/2}\exp\bigl(-\rho(2\tau
+\brm{p}^{2}/2)s\bigr).
\end{align}
One of the time-integrations and the $\brm{p}$-integration in
Eq.~(\ref{Dyson}) are easily done. Being interested in criticality we set
$\tau=0$. For simplicity, we restrict our attention to $\brm{r}=\brm{0}%
$. We expand the right hand side of Eq.~(\ref{hunger}) to first order in $L$
and $u$. To prepare for the renormalization step to follow, we mark
unrenormalized quantities with an open circle, $s\rightarrow\mathring{s}$,
$\rho\rightarrow\mathring{\rho}$ and so on as we do in the renormalization
schemes (\ref{renorScheme}) and (\ref{renorScheme2}). Expressing various
quantities through their renormalized counterparts as specified by
Eqs.~(\ref{renorScheme}) and (\ref{renorScheme2}) we arrive at
\begin{align}
&  \mathring{G}_{2}(\brm{0},t,\lambda)/\mathring{G}_{2}^{(0)}%
(\brm{0},t,\lambda)=1-\frac{u(2\rho\mu^{2}t)^{2-d\bot/2}}{8\Gamma
(1+\varepsilon/2)}\nonumber\\
&  \times\int_{0}^{1}dx\frac{(1-x)}{\bigl[x(1-x/2)\bigr]^{d\bot/2}%
}\Big\{1+x\rho t\bigl[2L(\lambda/2)-L(\lambda)\bigr]\Big\}\,.
\end{align}
The $x$-integral is performed using dimensional regularization. We obtain, up
to terms of $O(\varepsilon)$,
\begin{align}
&  \mathring{G}_{2}(\brm{0},t,\lambda)/\mathring{G}_{2}^{(0)}%
(\brm{0},t,\lambda)=1+\frac{u(2\rho\mu^{2}t)^{\varepsilon/2}}%
{8\Gamma(1+\varepsilon/2)}\nonumber\\
&  \times\Big[2+\rho t\bigl(2/\varepsilon+\ln2-1\bigr)\bigl[L(\lambda
)-2L(\lambda/2)\bigr]\Big]\,. \label{G2un}%
\end{align}
Equation~(\ref{renorScheme}) implies that the correlation function is
renormalized by
\begin{align}
G_{2}(\brm{0},t,\lambda)=Z^{-1}\mathring{G}_{2}(\brm{0},t,\lambda)\,.
\end{align}
Using again the renormalization factors stated in Eqs.~(\ref{renorFactors})
and (\ref{renorFactors2}) we get
\begin{align}
&  \mathring{\rho}\mathring{L}(\lambda)=\mathring{\rho}\mathring{w}_{r}%
\Lambda_{r}(\lambda)+\sum_{l}\mathring{\rho}\mathring{v}_{l}K_{l}%
(\lambda)\nonumber\\
&  =\rho\Big(Z^{-1}Z_{w_{r}}w_{r}\Lambda_{r}(\lambda)+\sum_{l}Z^{-1}Z_{v_{l}%
}v_{l}K_{l}(\lambda)\Big)\nonumber\\
&  =\rho\Big(L(\lambda)+\frac{u}{4\varepsilon}\bigl[L(\lambda)-2L(\lambda
/2)\bigr]\Big),
\end{align}
where we have used the homogeneity of the polynomials $\Lambda_{r}(\lambda)$
and $K_{l}(\lambda)$. From this relation we obtain after careful expansion in
$u$ and $L(\lambda)$ to first order%
\begin{align}
&  Z^{-1}\mathring{G}_{2}^{(0)}(\brm{0},t,\lambda)=\frac{(1-u/16)}%
{(4\pi\rho t)^{d_{\bot}/2}}\Big\{1\nonumber\\
&  -\rho t\Big[L(\lambda)+\frac{u}{4\varepsilon}\bigl[L(\lambda)-2L(\lambda
/2)\bigr]\Big]\Big\}\,.
\end{align}
Using this result in Eq.~(\ref{G2un}) we obtain after a final $\varepsilon
$-expansion%
\begin{align}
&  G_{2}(\brm{0},t,\lambda)=\frac{(1-u/16)}{(4\pi\rho t)^{d_{\bot}/2}%
}\Big\{\bigl(1+u/4\bigr)\bigl(1-\rho tL(\lambda)\bigr)\nonumber\\
&  +\rho t\frac{u}{8}(\ln2-1)\bigl[L(\lambda)-2L(\lambda/2)\bigr]\nonumber\\
&  +\rho t\frac{u}{4\varepsilon}\Big[\frac{(2\rho\mu^{2}t)^{\varepsilon/2}%
}{\Gamma(1+\varepsilon/2)}-1\Big]\bigl[L(\lambda)-2L(\lambda
/2)\bigr]\Big\}\nonumber\\
&  =\frac{(1+3u/16)}{(4\pi\rho t)^{2}}\Big\{1-\rho tL(\lambda)\nonumber\\
&  +\rho t\frac{u}{8}\bigl(\ln2-1+\ln(2\rho\mu^{2}t)+C_{E}%
\bigr)\bigl[L(\lambda)-2L(\lambda/2]\bigr)\Big\}\,. \label{End-Entw}%
\end{align}
Taking into account Eq.~(\ref{Wahl_X}) finally yields%
\begin{align}
&  G_{2}(\brm{0},t_{0},\lambda)=c_{0}\bigl(1+3u/16\bigr)\Big\{1-c_{1}%
\nonumber\\
&  \times\Big[\bigl(1+A_{r}u\bigr)w_{r}\Lambda_{r}(\lambda)+\sum
_{l}\bigl(1+A_{2l-1}u\bigr)v_{l}K_{l}(\lambda)\Big]\Big\} \label{End-Amp}%
\end{align}
with nonuniversal constants $c_{0}$ and $c_{1}$ and the amplitude $A_{r}$ as
stated in Eq.~(\ref{Ar}).



\begin{thebibliography}{99}
\bibitem {footnote_redundancy}The constant $\lambda_{M}$ which sets the width
of the voltage interval such that $[-\lambda_{M} \sqrt{M} < \lambda^{(\alpha)}
\leq\lambda_{M} \sqrt{M}]$ is a redundant scaling parameter in the limit
$D\rightarrow0$, $M\rightarrow\infty$ and hence our predictions are
independent of its value.
\end{thebibliography}
\end{document}